\documentclass{emulateapj}
\usepackage{apjfonts}
\usepackage{latexsym}
\usepackage{hyperref}
\shorttitle{SX Phoenicis in NGC 6541}
\shortauthors{Fiorentino et al.}
\begin{document}
\def\gsim{\;\lower.6ex\hbox{$\sim$}\kern-6.7pt\raise.4ex\hbox{$>$}\;}
\def\lsim{\;\lower.6ex\hbox{$\sim$}\kern-6.7pt\raise.4ex\hbox{$<$}\;}
\title{Blue straggler masses from pulsation properties. I. The case of NGC6541.}

\author{G. Fiorentino$^{1}$, B. Lanzoni$^{1}$, E. Dalessandro$^{1}$,
  F. R. Ferraro$^{1}$, G. Bono$^{2,3}$, M. Marconi$^{4}$}
\email{giuliana.fiorentino@oabo.inaf.it}
\affil{$^{1}$ Dipartimento di Fisica e Astronomia, Universit\`a degli
  Studi di Bologna, Viale Berti Pichat 6/2, 40127 Bologna, Italy}
\affil{$^{2}$Dipartimento di Fisica, Universit\'{a} di Roma Tor Vergata, Via della Ricerca Scientifica 1, 00133 Roma, Italy}
\affil{$^{3}$INAF-Osservatorio Astronomico di Roma, Via Frascati 33, 00040 Monte Porzio Catone, Italy}
\affil{$^{4}$INAF-Osservatorio Astronomico di Capodimonte, Via Moiariello 16, 80131 Napoli, Italy }

\begin{abstract}
We used high-spatial resolution images
acquired with the WFC3 on board HST to probe the population of
variable blue straggler stars in the central region of the poorly
studied Galactic globular cluster NGC 6541. The time sampling of the acquired multi wavelength (F390W, F555W and F814W)
data allowed us to discover three WUma stars and nine SX
Phoenicis. Periods, mean magnitudes and pulsation modes have been
derived for the nine SX Phoenicis and their masses have been estimated
by using pulsation equations obtained from linear non adiabatic models. We found masses in the range 1.0-1.1M$_\odot$, with
an average value of $1.06 \pm 0.09$ M$_\odot$ ($\sigma =0.04$),
significantly in excess of the cluster Main Sequence Turn Off mass
($\sim0.75$M$_\odot$). A mild trend between mass and luminosity seems also
to be present. The computed pulsation masses turn out to be in very
good agreement with the predictions of evolutionary tracks for single
stars, indicating values in the range $\sim 1.0-1.2$ M$_\odot$ for most of the
BSS population, in agreement with what discussed in a number of previous
studies.
\end{abstract}


\section{Introduction}
\label{intro}
Dense stellar environments like Galactic globular clusters (GCs) are
populated by a large variety of {\it exotica}, including Blue
Straggler Stars (BSSs), X-ray binaries, millisecond pulsars and
cataclysmic variables
\citep[see][]{bailyn95,pooley06,paresce92,bellazzini95,ferraro01,ransom05,freire08}. Among
them, BSSs certainly are the most numerous. They were discovered by
\citet{sandage53} in the outer regions of the Galactic GC M3 and,
since then, they have been detected 
not only in GCs \citep[see][]{piotto04,leigh11}, but also in
open clusters \citep{mathieu09}, and dwarf galaxies
\citep{mateo95,mapelli07,monelli12a}.

In the optical color magnitude diagram (CMD) BSSs are bluer and
brighter than the main sequence (MS) turn-off (TO) stars, defining a
sequence that typically spans more than 2-2.5 magnitudes above
the cluster TO. Hence they mimic a younger stellar population with
masses larger than normal cluster stars.  Indeed, masses of
the order of M$=(1.0 - 1.7) $M$_\odot$ have been estimated by direct
measurements, although with large uncertainties
\citep{shara97,gilliland98,demarco05}.\par

The origin of BSSs is still a matter of debate. They could be
originated from collision-induced mergers, most likely in dense
stellar environments \citep{hills76,leonard89}, or by mass exchange in
primordial binary systems
\citep{mccrea64,zinn76,knigge09,ferraro06a,ferraro06b}. These two
scenarios can possibly co-exist within the same cluster
\citep{ferraro95,ferraro09b}. Independently of their formation
mechanism, BSSs are the brightest among the most massive stars in the
host cluster and therefore are ideal tools to probe the dynamical
evolution of stellar systems. Indeed, the observed shape of their
radial distribution within the cluster has been recently used to
define the so-called {\it "dynamical clock"} \citep{ferraro12}, able
to rank GCs on the basis of the dynamical stage reached. Since the
engine of such a clock is dynamical friction and since the dynamical
friction efficiency directly depends on the object mass, an accurate
determination of BSS masses is of paramount importance for the
calibration of such a clock.\par

\begin{figure*} 
\includegraphics[width=8.5cm]{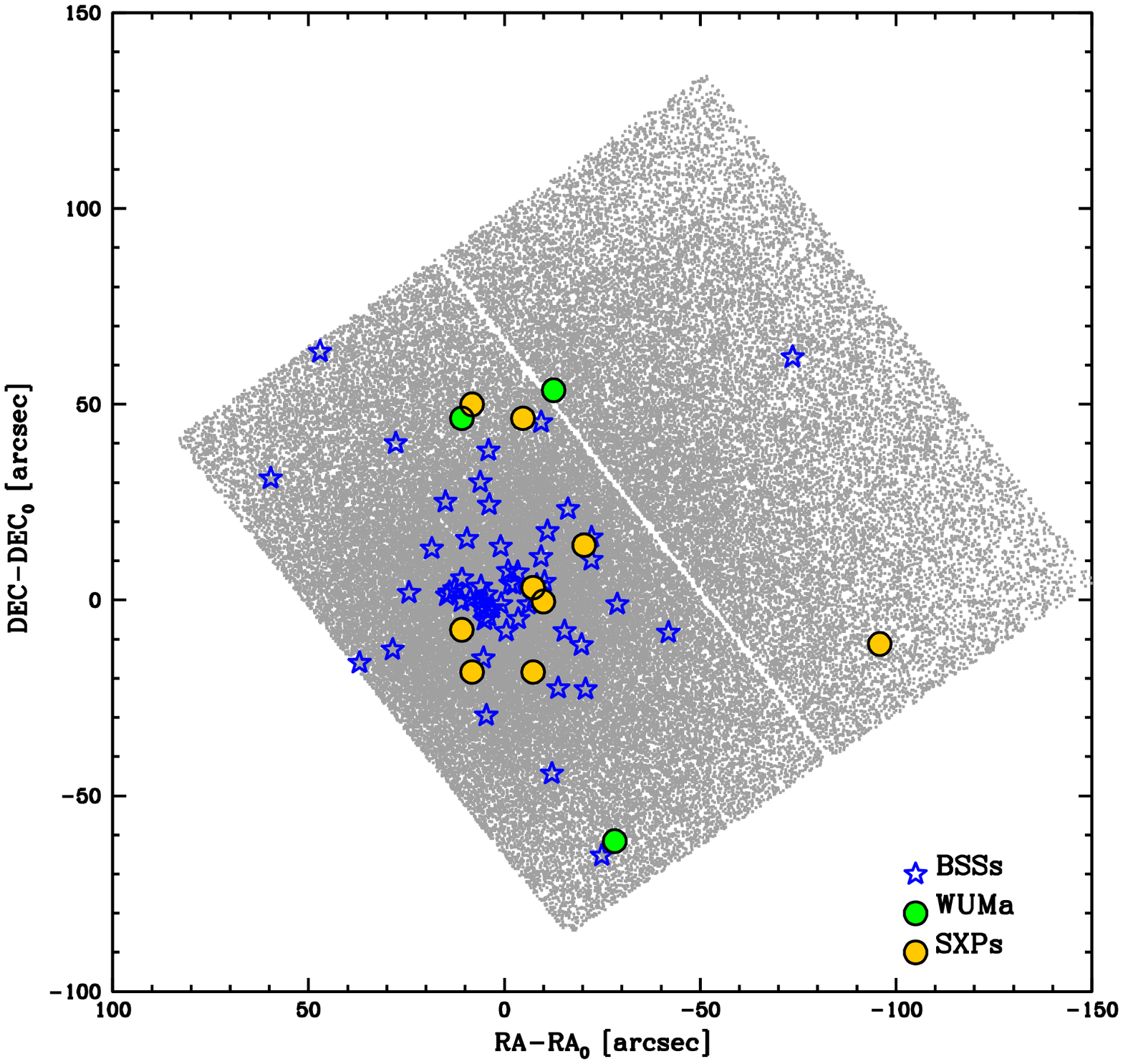} 
\includegraphics[width=8.5cm]{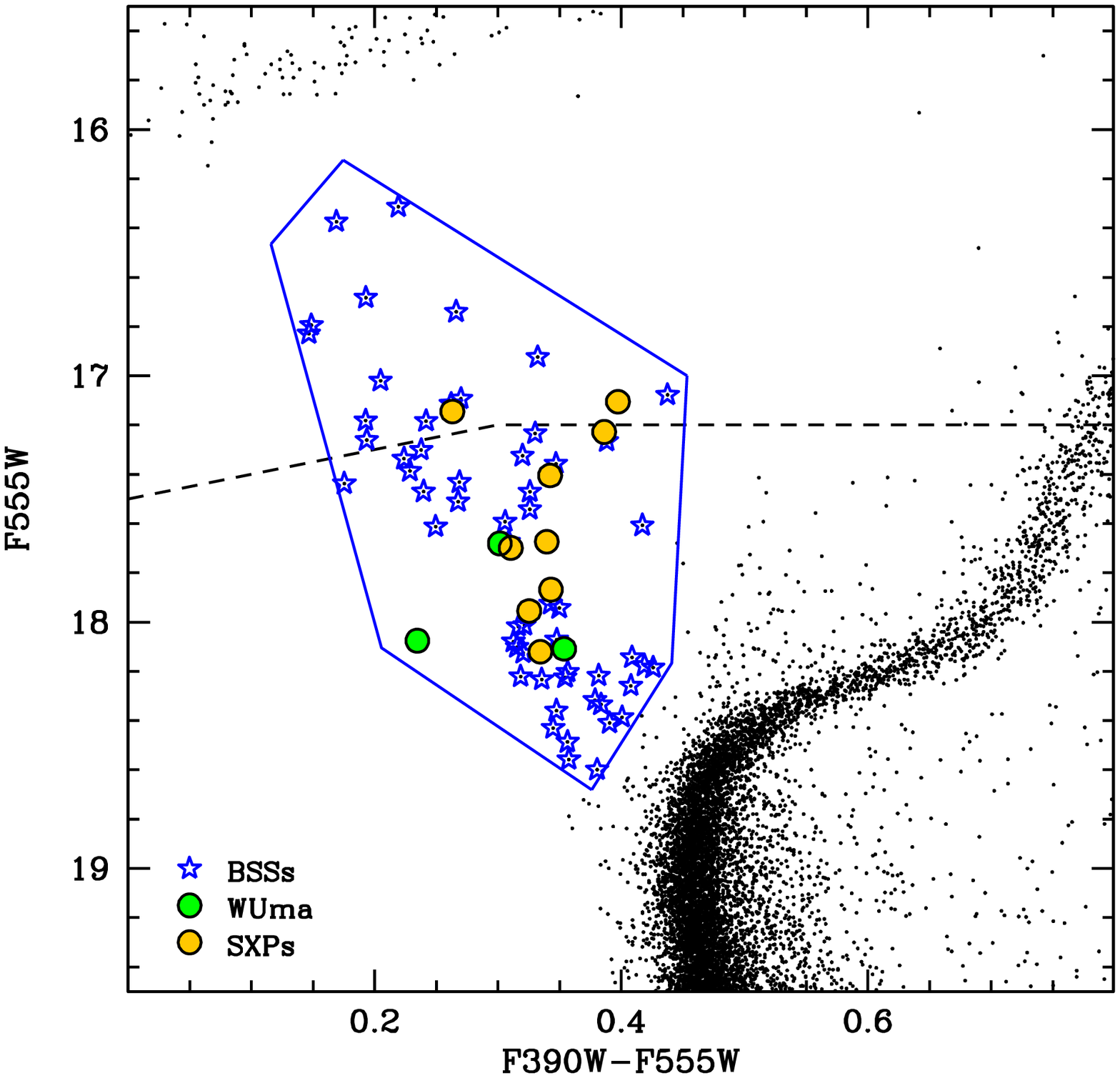} 
\caption{ {\it left} -- Map of the WFC3 data. The location of the selected BSSs is
  highlighted: blue starred symbols mark non-variable BSSs, green and
  orange circles indicate WUma and SXP BSSs, respectively.  Most of
  the selected BSSs are located in Chip\#1, where the cluster center
  is also positioned.  {\it right} -- (F555W, F390W$-$F555W) CMD of NGC 6541 zoomed in the BSS
  region.  The box used to select the BSS population is shown with
  blue contours. Non variable BSSs are marked as blue starred symbols,
  while variable BSSs are highlighted with green (WUma) and orange
  (SXP) filled circles.  The saturation limit of our dataset is also
  marked for reference (dashed line).
\label{fig:fccmd}} 
\end{figure*}

Very interestingly, BSSs cross the faint extension of the classical
Instability Strip (IS), where they are observed as SX Phoenicis stars \citep[SXPs, see][]{pych01,jeon03,jeon04,mcnamara11,arellano11}.  These objects show a photometric variability on very short
periods (P$\lsim$ 0.1 d) and can be unstable for radial and non-radial
pulsation. Hence, their pulsation properties can be used to estimate
their masses. The Petersen's diagram (period ratio vs
the longer period; \citealp{petersen78,stellingwerf78,gilliland98}),
which is largely used for variable stars pulsating simultaneously in
different modes, is very little sensitive to the mass in the period range
typical of SXPs. However these variables follow the classical pulsation equation, relating the observed period to the intrinsic stellar
parameters such as mass, luminosity and effective temperature P(M, L, T$_{\rm
  eff}$), for any
given pulsation mode and chemical composition. This relation can
indeed be used to estimate the star mass, once the full set of
variables is observed and/or assumed
\citep{stellingwerf79,dacosta86,marconi04,caputo05,fiorentino06}.  For
a precise determination of the pulsation equations, a complete theoretical scenario using a nonlinear and non adiabatic
approach, also accounting for the convection, should be defined
\citep{bono97a,fiorentino02,marconi05}. This would be able to fully
describe the pulsation properties of SXPs including their amplitudes
and the red (cold) boundary of the IS. However, such a theoretical
framework is still missing due to the huge amount of computation time
necessary to integrate conservation equations for these high gravity
pulsators. Nevertheless, theoretical models based on a radial, linear,
non adiabatic approach to the stellar pulsation have been shown to
reproduce quite confidently the SXP periods and the blue (hot)
boundary of the IS \citep{stellingwerf79,gilliland98,santolamazza01},
thus supporting the use of these approximated pulsation equations for
the mass estimate.  Attempts to estimate SXP masses have been performed
using the pulsation equation based on \citet{jorgensen84} linear
theoretical models, and masses of about 1.5 M$_\odot$ up to twice the
main sequence TO mass have been obtained for a sample of BSSs in NGC
5053 \citep{nemec95}.

In the present paper we use linear and non adiabatic theoretical
pulsation models from \citet{santolamazza01} to estimate the
mass of a sample of BSSs pulsating as SXPs in the Galactic globular
cluster (GC) NGC 6541. This is an old \citep[13.25$\pm$1
Gyr][]{dotter10} and metal-poor ([Fe/H]$=-1.76 \pm
-0.02$, \citealp{lee02}) GC with an extended blue Horizonthal Branch
\citep{dotter10}. It is located at 3 kpc from the Galactic Center and classified as a
post-core collapse cluster \citep{harris96}.
Here new HST/WFC3 data are used to detect variable BSSs in the central region of NGC 6541. We found a sizeable sample of
this exotic population and we discuss the
mass determination for nine SXP pulsating BSSs.

The paper is organized as follows. The HST dataset and the data
reduction procedure are described in Sec. 2. The BSS selection and
their short-period variability study are described in Sec. 3. In
Sec. 4 we present the method adopted to discriminate the pulsation
mode of the nine selected SXPs, whereas in Sec. 6 we derive their
mass. A discussion Section closes the paper.

\section{Observations and data reduction}

The data-set presented in this paper has been acquired using the UVIS
channel of the HST/WFC3 on 2012 February 24 (PI: F. R. Ferraro, GO:
12516). The cluster is centered in chip\#1 (see
Fig.~\ref{fig:fccmd}, left panel). The data set consists of twelve F390W, ten F555W
(narrow $V$) and thirteen F814W (wide $I$) frames. Each frame has been
integrated for an exposure time of 348 s, 150 s and 348 s,
respectively.  The observing strategy includes a small dithering of a few
pixels to correct CCD blemishes, artifacts and false detections. The
three different filters were alternated during each orbit to optimize
the final total exposure time. 

The data reduction has been performed using DAOPHOT/ALLSTAR/ALLFRAME
\citep{stetson87,stetson94} on flat-fielded and bias subtracted
(\_flt) images once the pixel area map corrections were accounted
for. The Point Spread Functions (PSFs) were derived on chip\#1 and \#2
separately using the brightest, isolated and non saturated
stars. Typically, more than 600 stars were selected. These PSFs were
used to perform ALLSTAR photometry in each frame. The obtained
individual star lists have been combined using DAOMATCH in a single
master list to run ALLFRAME. In this master list only stars appearing
in at least 20 out of 35 images were included. Finally, the ALLFRAME
output star lists were combined using DAOMASTER for each filter. For
each star, we have homogenized the different magnitudes
\citep[e.g.,][]{ferraro91,ferraro92} to obtain their mean magnitudes
and the corresponding time-series.  The final list includes 72,616
stars.

In order to calibrate instrumental F390W, F555W and F814W magnitudes
in the VEGAMAG photometric system we have followed the procedure
described in \citet{dalessandro13c}, which includes
Charge Transfer Efficiency correction for the WFC3 camera.  A portion
of the resulting (F555W, F390W$-$F555W) CMD, zoomed in the BSS region,
is shown in Figure \ref{fig:fccmd} (right panel). By following the same paper and
according to the estimate of the UVIS channel performance, we have
also computed the saturation threshold of our photometry. Following
the instructions of the WFC3 Data
Handbook, we used 72000 e$^-$ as the value of the central pixel
saturation and we adopted an aperture of 1$\times$1 pixel for the encircled
energy in the three filters: EE(F814W)$= 0.149$, EE(F555W)$=0.184$ and
EE(F390W)$=0.180$. Then the flux close to saturation has been
determined and transformed into magnitudes by applying the same
equations used to calibrate our photometry in the HST VEGAMAG
photometric system.  The resulting saturation threshold is marked in
Fig. \ref{fig:fccmd} (right panel, dashed line).

Finally, to obtain a precise absolute astrometric solution we first
accounted for geometric distortion following the prescriptions
discussed in the WFC3 Data Handbook.  Then, the instrumental positions
of our stars have been cross correlated with a wide field catalog
(P. B. Stetson, private communication) previously reported to the
absolute astrometric system using the stars in common with the Guide
Source Catalogue 2.3 (GSC2.3), by means of CataXcorr
\citep[see][]{dalessandro11}.  Both the WFC3 and the wide field
catalogs will be used in a forthcoming paper to properly determine the
center and density profile of NGC 6541 and to constrain the mass and
luminosity functions of this cluster.

\section{Blue Straggler Star selection and variability search}
\label{lc}
The BSS population has been selected on the basis of the star location in the CMD (as objects bluer and brighter than the MS-TO), with conservative colour and
magnitude limits aimed at reducing at most the contamination from stellar blends
near the MS-TO and the sub-giant branch. In addition, a simultaneous check of the
(F555W, F390W-F555W) and the (F555W, F555W-F814W) planes has been
performed, and only stars located in the ``appropriate'' region in
both CMDs have been selected as fiducial BSSs, following the general
approach routinely adopted in previous papers
\citep{ferraro06a,dalessandro08}. The resulting shape of the BSS
selection box in the (F555W, F390W-F555W) plane is shown with blue
contours in the right panel of Fig.~\ref{fig:fccmd}.
The final sample consists of 70 BSSs spanning $\sim 2$ mag in
the F555W pass band and approaching the Horizontal Branch luminosity.
The spatial distribution of these stars within NGC 6541 is shown in
Fig.\ref{fig:fccmd} (left panel), revealing that they are well
segregated in the cluster center \citep[as quoted by][2010
  version]{harris96}.

As apparent from Fig. \ref{fig:fccmd} (right panel), sixteen stars
fall above the saturation threshold.  However, all variable BSSs
(green and orange circles; see below), which are the main targets of
this study, are either below or very close to the saturation
limit. Hence they are marginally affected by the photometric
saturation. It is worth mentioning that, while possible variability
signals may be hidden by saturation effects at birghter magnitudes,
the most luminous BSSs are so blue that they unlikely cross the
pulsation IS.

Our data sampling was collected within 13 HST orbits and consists of 35 images
over a time interval of about 7 hours. However, thanks to the high
photometric accuracy possible with HST (with errors $\sigma_{\rm F390W}-\sigma_{\rm F555W}-\sigma_{\rm F814W} \lsim
0.015$ mag at the BSS magnitude levels), it has been possible to perform a
variability study and search for variable BSSs with short periods and
relatively large amplitudes (e.g., A$_{\rm F555W} \gsim$ 0.05 mag), as
already done for the case of NGC 362 \citep{dalessandro13c}.

To this end, we have performed a Fourier time series analysis for
  all the selected BSSs  in the three filters simultaneously by using
  GRATIS \citep[GRaphical Analyzer of TIme Series, developed by
  P. Montegriffo at the INAF-Bologna Observatory; see][and reference
  therein for details]{fiorentino10a,fiorentino10b}. In particular, we
  have used low order Fourier coefficients to match the light curves
  in each pass band as a function of the phased Heliocentric Julian
  date. Only those stars showing variability with the same period
  within 1\% in all the three filters were considered as {\it bona
    fide} variables.  In particular, out of the 70 selected BSSs, nine
  stars  with period smaller than 0.1 days have been identified as
  SXPs and  three stars with period longer than 0.1 days where
  classified as  WUMa stars. It is worth noticing that the detected
  sample of variable stars is far to be complete.

\begin{figure} 
\includegraphics[width=8.5cm]{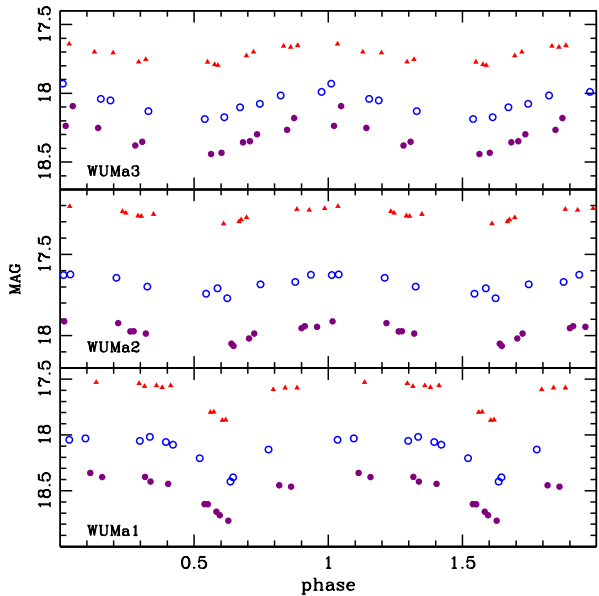} 
\caption{Phased light curves for the WUMa stars detected within our
  BSS sample. Different filters are color coded in each panel: purple
  light curve is for the F390W pass band, blue for the F555W and red
  for the F814W.\label{fig:lc_wuma}} 
\end{figure}

\begin{figure} 
\includegraphics[width=8.5cm]{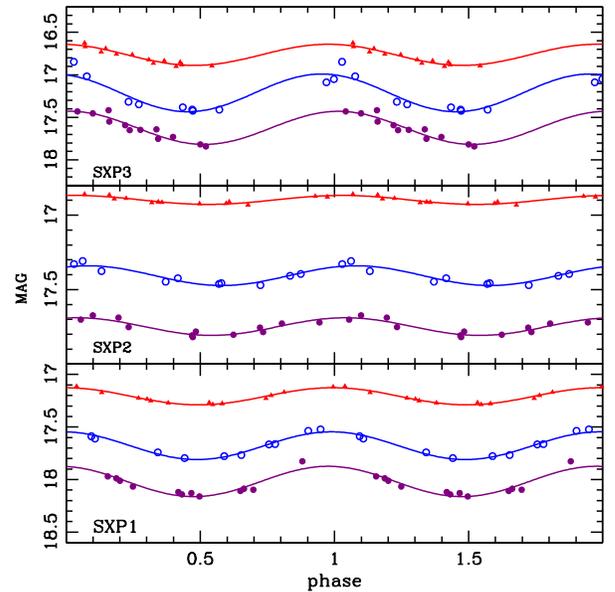} 
\caption{Phased light curves for three SXPs detected in our BSS
  sample. The color code is as in Fig. \ref{fig:lc_wuma}.
\label{fig:lc_sxp1}} 
\end{figure}

\begin{figure} 
\includegraphics[width=8.5cm]{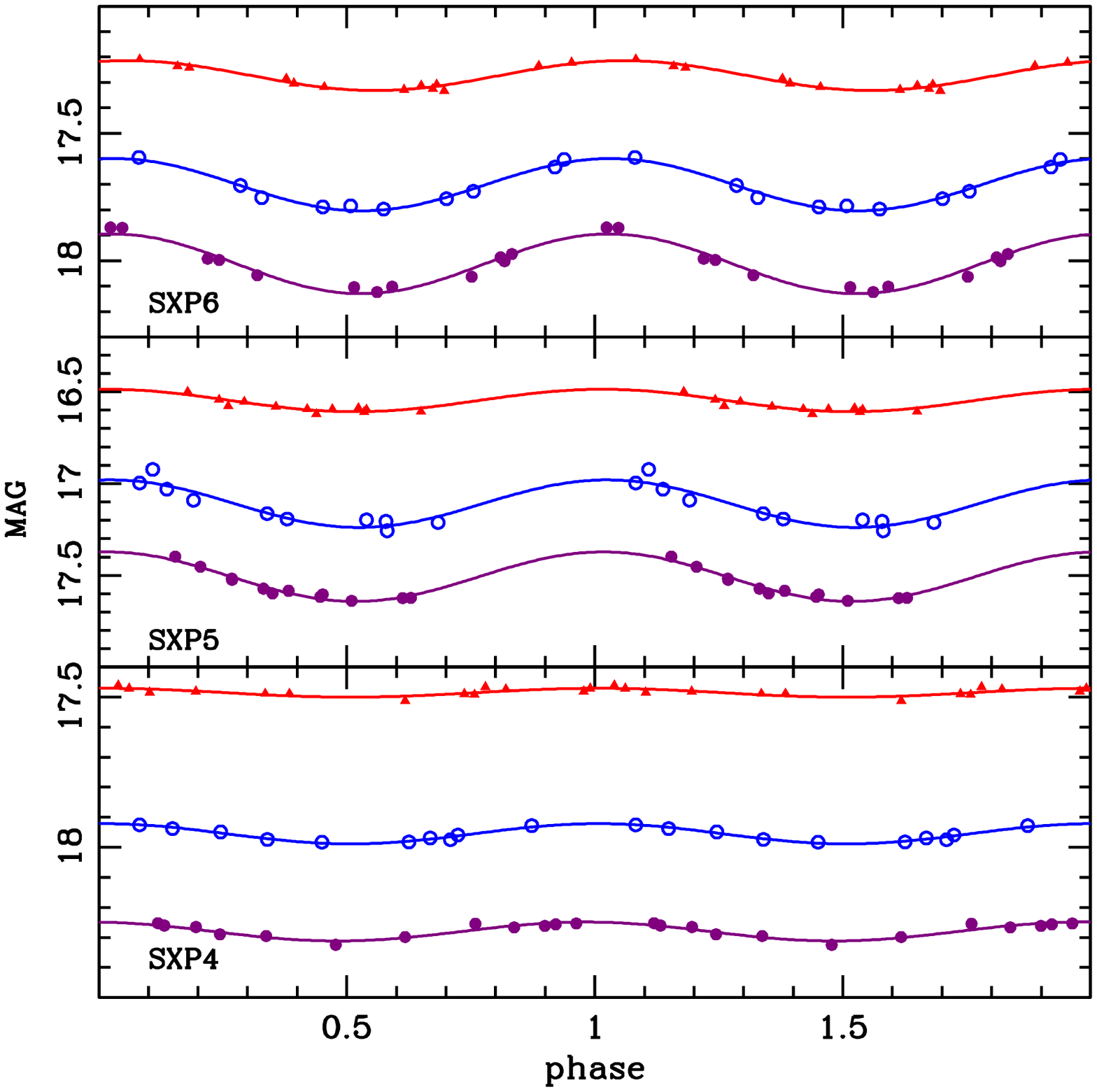} 
\caption{Same as in Fig.~\ref{fig:lc_sxp1} for three other SXP BSSs.
\label{fig:lc_sxp2}} 
\end{figure}

\begin{figure} 
\includegraphics[width=8.5cm]{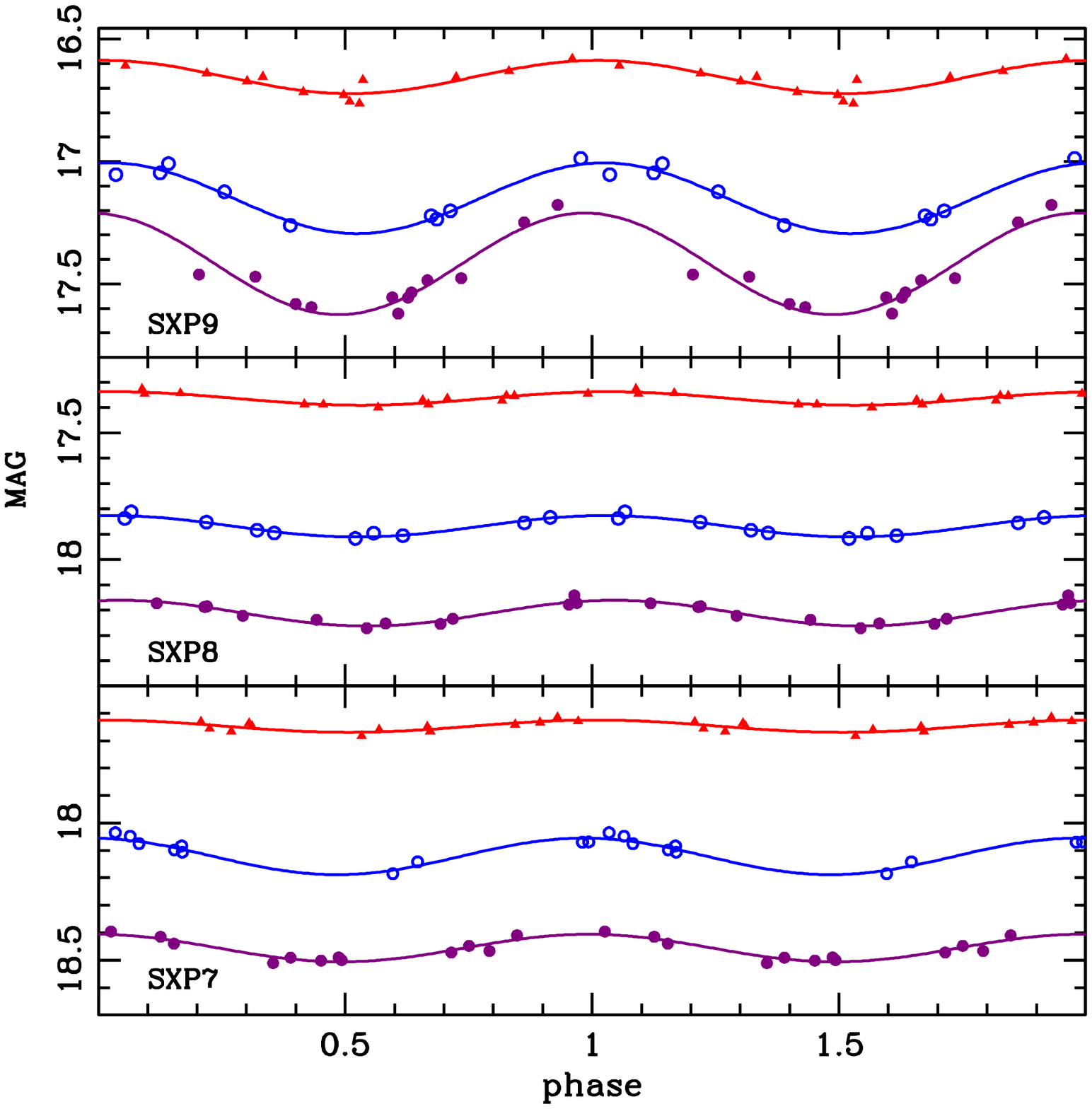} 
\caption{Same as in Fig.~\ref{fig:lc_sxp1} for three other SXP BSSs.  
\label{fig:lc_sxp3}} 
\end{figure}

The phased light curves of the three WUma stars are shown in Fig.~\ref{fig:lc_wuma} for
the three filters. We remember here that WUMa stars are eclipsing binary systems. WUMa1 shows a not very sinusoidal behavior, with a
flattening of the light curve at bright magnitudes due to the
contribution of both the companions. The remaining two WUMa
stars show, instead, a sinusoidal light curve typical of two very
close components in a binary system. Our sample of SXPs shows sinusoidal light
curves (see Figures \ref{fig:lc_sxp1}-\ref{fig:lc_sxp3}), as expected
in the case of an almost adiabatic behavior with large amplitudes.
Despite the modest sampling of our data, we stress here that our
  classification stars as SXPs is fully supported by their pulsation
  properties, e.g. they follows a very tight and well defined PL
  relation (see Section~\ref{mode}).

Their amplitudes range from 0.06 to 0.39 mag in the F555W filter,
and their periods varies from 0.032 to 0.065 days.  We stress here
that, due to both the photometric errors and the limited temporal
sampling of our data, only high amplitude variations can be detected,
and our SXP sample is therefore far to be complete. In addition, it is
not possible to firmly assess the existence of more than one frequency, and
in turn mixed mode pulsators can not be solidly recognized.

We derived the mean magnitudes of the detected variable stars using
the truncated Fourier series that best match the data. The mean
instrumental magnitudes were then calibrated using the same method
adopted for non variable stars and described in the previous Section.
The intensity-averaged mean magnitudes, periods, amplitudes and epochs (time at the
maximum of the light curve) are listed in Table \ref{tab1}, together
with the star coordinates.  The position in the CMD of the detected
SXP and WUMa stars, plotted according to the values given in Table
\ref{tab1} are shown in Fig. \ref{fig:fccmd} (right panel).

\begin{center}
\begin{deluxetable*}{lccccccccccc}
 \tabletypesize{\tiny}    
\tablecaption{Parameters of the variable BSSs detected in NGC 6541. \label{tab1}}
\tablehead{\colhead{ID}&\colhead{$\alpha_{J2000.0}$}&
\colhead{$\delta_{J2000.0}$}&\colhead{Period}&\colhead{Epoch}&
\colhead{A$_{\rm F390W}$}& \colhead{A$_{\rm F555W}$} &\colhead{A$_{\rm F814W}$}&
\colhead{<F390W>}& \colhead{<F555W>} &\colhead{<F814W>}&\colhead{<M/M$_{\odot}$>}\\ 
\colhead{}&\colhead{deg}&
\colhead{deg}&\colhead{days}&\colhead{(-2455900)}&
\colhead{mag}& \colhead{mag} &\colhead{mag} &
\colhead{mag}& \colhead{mag} &\colhead{mag}&\colhead{}}
\centering
\startdata
WUMa1     & 272.014  & -43.702 & 0.26$\pm$0.04  &80.850 &0.45  & 0.45  &0.34& 18.46     & 18.11    &17.62 & ''\\
WUMa2     & 271.999  & -43.732 & 0.20$\pm$0.02  &81.450 &0.14  & 0.12  &0.09 &  17.98     & 17.68    &17.26&''\\
WUMa3    & 272.005  & -43.700 & $\sim$0.44$\pm$0.10$^a$  &80.850  &0.29  & 0.23  &0.16 &  18.31     & 18.08    &17.72&''\\
SXP1$^{F}$  & 272.002  & -43.711 & 0.050$\pm$0.001  &80.970  &0.29  & 0.26  &0.16 &  18.01     & 17.67    &17.21  & 1.03$\pm$0.15\\
SXP2$^{FO}$ & 272.013  & -43.701 & 0.0461$\pm$0.003&80.959  & $\gsim$0.12  & 0.13  &0.06 &  17.75     & 17.40    &16.90& 1.27$\pm$0.18$^{F}$-1.07$\pm$0.16$^{FO}$\\
SXP3$^{F}$  & 272.007  & -43.714 & 0.0649$\pm$0.004&80.924  &$\gsim$  0.39  & 0.44  &0.25 &  17.61     & 17.20    &16.76& 1.13$\pm$0.17\\
SXP4$^{F}$  & 272.006  & -43.715 & 0.041$\pm$0.003 &80.956  &$\gsim$0.06  & 0.07  &0.03 &  18.28     & 17.95    &17.49& 1.00$\pm$0.15\\
SXP5$^{F/FO}$ & 272.007  & -43.720 & 0.065$\pm$0.002 &80.980  &0.27  & 0.26  &0.12 &  17.50     & 17.11    &16.55& 1.23$\pm$0.18$^{F}$-1.06$\pm$0.16$^{FO}$\\
SXP6$^{F}$  & 272.014  & -43.717 & 0.0425$\pm$0.001&81.008  &0.23  & 0.20  &0.12 &  18.01     & 17.70    &17.27& 1.11$\pm$0.16\\
SXP7$^{F}$  & 272.013  & -43.720 & 0.032$\pm$0.001 &80.970  &$\gsim$0.10  & 0.13  &0.04 &  18.46     & 18.12    &17.65& 1.05$\pm$0.15\\
SXP8$^{F}$  & 272.008  & -43.702 & 0.046$\pm$0.002 &80.985  &0.10  & 0.08  &0.05 &  18.21     & 17.87    &17.36& 0.98$\pm$0.15\\
SXP9$^{FO}$ & 271.973  & -43.718 & 0.059$\pm$0.003 &81.015  &0.42  & 0.29  &0.15 &  17.41     & 17.14    &16.66  & 1.27$\pm$0.19$^{F}$-1.09$\pm$0.16$^{FO}$ 
\enddata
\tablecomments{Name, coordinates, period, epoch, amplitude, mean
  magnitude for the WUma and SXP stars identified in the BSS sample of
  NGC 6541. For the nine SXPs, also the likely pulsation mode
  ($^{F}$/$^{FO}$) and the estimated pulsation mass are indicated
    (the latter corresponds to the average among the values obtained
    in the three filters; see Sect. \ref{sec:mass}).  For those stars
    with uncertain mode classification, we have listed two possible
    values for the mass as computed using F or FO PL relations. $^a$
 This is the best period that match our data in the range 0.01--0.9
 days. However, we note that our time sampling covers only 0.3 days, thus this value may be very uncertain.}
\end{deluxetable*}
\end{center}

\section{Pulsation mode identification}
\label{mode}

\begin{figure} 
\includegraphics[width=8.5cm]{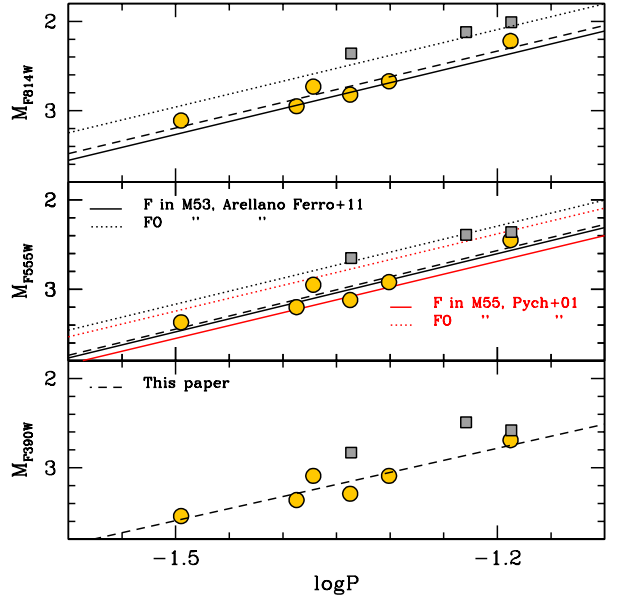} 
\caption{Location of the detected SXP BSSs (orange circles) in the
  Period-Absolute Magnitude plane, for the three available pass bands under the assumption of a reddening E$(B-V)=0.14$
mag and a distance modulus $\mu_0 =14.29$ mag
\citep[see][]{lee02}. Orange
  circles and grey squares represent F and FO
  pulsators respectively. The
  solid lines represent the PL relations derived for the Fundamental
  (F) mode pulsators observed in M53 (black; \citealp{arellano11}) and
  in M55 (red; \citealp{pych01}). The dotted lines correspond to these
  relations shifted assuming P$_{\rm FO}$/P$_{\rm F}=0.783$ and they
  mark the loci of the First Overtone (FO) pulsators. The dashed lines
  represent the best-fits to our data, derived by using only the
  likely F pulsators (see Section \ref{mode} for details).
\label{fig:pl}} 
\end{figure}

\begin{center}
\begin{deluxetable}{lccc}
 \tabletypesize{\scriptsize} 
\tablecaption{PL-relations for Fundamental mode pulsators.\label{tab2}}
\tablewidth{0pt}
\tablecolumns{6}
\tablehead{\colhead{FILTER}&\colhead{$\alpha$}&\colhead{$\beta$}&\colhead{$\sigma$}}
\startdata
M$_{F390W}$  & $-0.46$   & $-2.70\pm 0.51$ & 0.10 \\
M$_{F555W}$  & $-0.95$   & $-2.93\pm 0.53$ & 0.11 \\
M$_{F814W}$  & $-1.11$   & $-2.87\pm 0.42$ & 0.09 
\enddata
\tablecomments{Numerical coefficients of the derived PL-relations
  expressed as MAG $= \alpha + \beta\times\log$ P. The last
    column quotes the rms scatter of each relation.}
\end{deluxetable}
\end{center}

In order to estimate the star mass from the pulsation equation, the
first step is to determine the pulsation mode of the investigated
variables.  For well known pulsating stars crossing the IS, such as RR
Lyrae, the first overtone (FO) mode pulsators
show a more sinusoidal light curve with respect to the fundamental (F)
mode pulsators. Moreover, the FO amplitudes are typically smaller than
the F ones. Hence, the pulsation mode can be easily estimated from the
morphology of the light curve. In those cases where these general
rules are not easily applied, as for SXPs stars and Anomalous
Cepheids, more detailed pulsation properties can also be adopted; e.g., the location in the period luminosity
(PL) plane when variable stars are located at the same distance \citep{marconi04,fiorentino06}. We note that in the recent
years, the SXP PL relation has been accurately investigated in several
Galactic GCs \citep[e.g.][and reference
  therein]{pych01,jeon03,jeon04,mcnamara11,arellano11} and in dwarf
galaxies \citep[e.g.  LMC, Fornax, Carina, see][]{vivas13} to
determine the distance of the host stellar system. Here, instead, we
use it only to determine the pulsation mode of our SXP sample, which
is needed to estimate the SXP pulsation mass (see Section
\ref{sec:mass}).

Empirical and theoretical results suggest significantly different
periods for the F and the FO mode pulsations in SXPs, with a ratio
P$_{\rm FO}$/P$_{\rm F} \sim 0.783$
\citep{santolamazza01,jeon03,arellano11,pych01}. Thus the two
pulsation modes can be confidently distinguished in the
Period-Absolute Magnitude plane. Indeed, abundant populations of SXPs
in metal-poor GCs (21 SXPs with [Fe/H]$=-2.10$ in M53, and 24 SXPs at
[Fe/H]$=-1.94$ in M55; \citealp{arellano11, pych01}, respectively)
define two separate sequences well fitted, respectively, by the
PL-relation derived for F mode pulsators (solid lines in Figure
\ref{fig:pl}) and by the same relation shifted under the assumption
P$_{\rm FO}$/P$_{\rm F}= 0.783$ (dotted lines).\footnote{These
  relations have been obtained in the Johnson Kron--Cousins photometric
  system, we do not expect any large difference when using F555W and
  F814W magnitudes as in the case of classical Cepheids
  \citep{fiorentino13a}.} We therefore used these relations to
infer the pulsation mode of our SXPs, by assuming a
reddening E$(B-V)=0.14$ mag \citep{cardelli89} and a distance modulus $\mu_0 =14.29$ mag
\citep{lee02} for NGC6541.

As shown in Fig. \ref{fig:pl}, most of our variables follow the F mode
PL relation in both the F555W ($V$) and the F814W ($I$) bands, with
the possible exception of three stars, namely SXP2, SXP5 and SXP9. 
SXP2 and SXP9 seem to well follow the PL relations for FO mode
  pulsators in all the three filters. The case of SXP5 is more
  ambiguous being this star located close to the F mode PL relation in
  F390W and F555W passbands whereas it follows better the FO mode PL
  relation in the F814W filter. Here it is worth noticing that, since
  SXP5 has a quite large amplitude and its light curve is not fully covered by our data, its mean magnitude estimates
may be not very accurate. For this reason, in the following discussion
we will consider that SXP5 can be classified either as F or FO
pulsator.

Finally, we found a quite good agreement (in terms of both the slopes and the zero
points) between the linear fit to our likely F pulsators (dashed lines
in Fig. \ref{fig:pl} and Table \ref{tab2}) and the relations quoted by
\citet{pych01} and \citet{arellano11}. Hence, we can reasonably use our
data to determine, for the first time, the PL relation of SXPs in the
F390W-band (see bottom panel of Fig. \ref{fig:pl} and Table
\ref{tab2}).

\section{Pulsation masses}
\label{sec:mass}

\begin{figure} 
\includegraphics[width=8.5cm]{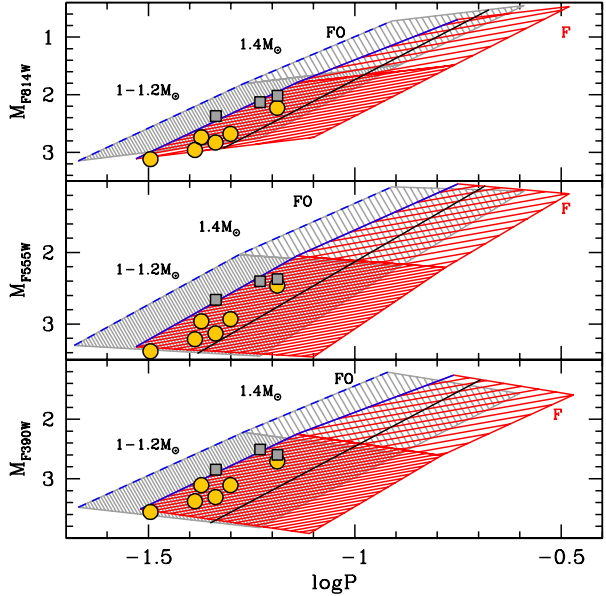} 
\caption{Grids of linear non adiabatic models
  \citep[from][]{santolamazza01} in the Period-Absolute Magnitude
  plane for the available set of filters. Red and grey colors
  correspond to models computed for F and FO mode pulsators,
  respectively.  Narrow and large shadings correspond to different
  masses: 1.0-1.2 M$_\odot$ and 1.4 M$_\odot$ respectively. The  blue lines represent the theoretical Fundamental (solid) and First-Overtone (dashed) Blue Edges of the IS. The black
  solid lines mark the empirical red edge of the IS adopted ``ad hoc''
  for deriving the theoretical PL relations, see the text for details. SXPs 
  are shown for comparison with the same symbols as in
  Fig.~\ref{fig:pl}.
\label{fig:is}} 
\end{figure}

\begin{center}
\begin{deluxetable}{lccccc}
 \tabletypesize{\scriptsize}    
\tablecaption{Pulsation equations used to estimate the star mass. \label{tab3}}
\tablewidth{0pt}
\tablehead{\colhead{FILTER}&\colhead{$\alpha$}&\colhead{$\beta$}&\colhead{$\gamma$}&\colhead{$\delta$}&\colhead{$\sigma$}}
\startdata
\multicolumn{6}{c}{F mode}\\ 
\hline
M$_{F390W}$  & $0.13$  & $-0.18\pm 0.04$ & $-0.38\pm 0.13$ & $0.02\pm 0.01$ & 0.04 \\
M$_{F555W}$  & $0.05$  & $-0.20\pm 0.04$ & $-0.46\pm 0.13$ & $0.01\pm 0.01$ & 0.04 \\
M$_{F814W}$  & $-0.27$ & $-0.33\pm 0.05$ & $-0.89\pm 0.15$ & $0.01\pm 0.01$ & 0.03 \\
\hline
\multicolumn{6}{c}{FO mode}\\ 
\hline
M$_{F390W}$  & $0.13$  & $-0.14\pm 0.03$ & $-0.26\pm 0.08$ & $0.01\pm 0.01$ & 0.04 \\
M$_{F555W}$  & $0.04$  & $-0.17\pm 0.03$ & $-0.34\pm 0.09$ & $0.01\pm 0.01$ & 0.04 \\
M$_{F814W}$  & $-0.31$ & $-0.29\pm 0.03$ & $-0.77\pm 0.11$ & $0.01\pm 0.01$ & 0.03 
\enddata
\tablecomments{Numerical coefficients of the pulsation equations
  derived from linear non adiabatic models and expressed as $\log$
  M/M$_\odot$ = $\alpha + \beta \times $MAG$+\gamma \log$ P $+\delta
  \log$ Z/Z$_\odot$.}
\end{deluxetable}
\end{center}

Once the pulsation modes of our SXP sample have been determined
(Sect. \ref{mode}), we estimate the BSS mass through pulsation
equations P(M, L, T$_{\rm eff}$) derived from the linear and non
adiabatic pulsation models of \citet{santolamazza01} for the first two
modes. In that paper, the authors present a grid of models assuming
three values for the star mass (1.0, 1.2 and 1.4 M$_\odot$) and two
different metallicities ([Fe/H]$=-2.2$ and $-1.3$ dex) that bracket
the iron abundance of NGC 6541 ([Fe/H]$=-1.76$; \citealp{lee02}). For
each mass, several luminosity levels are adopted. The models also span
a large range in effective temperatures to reproduce the extension of
the IS. However, the major limit of the adopted models is that they do
not account for the efficiency of the convection flux during the
pulsation. This becomes especially important at cold effective
temperatures (T$_{\rm eff} \sim 6000 K$ for the SXPs mass range), where convection can balance the temperature gradient in the
stellar envelope, with the final effect of quenching the stellar
pulsation. This means that the linear non adiabatic approach is unable
to predict the location of the red boundary of the IS and the explored
temperatures cover a large range arbitrarily fixed by the
authors. 
To overcome this limitation, we decide to adopt the \emph{observed}
red boundary of the IS, which seems to be a quite reasonable
assumption.

In order to compare the models with our data we have transformed the
theoretical luminosities and effective temperatures into the VEGAMAG
HST photometric system, as described in \citet[][]{fiorentino13a}. The
region covered by the grid of F and FO models in the PL plane is shown
in Fig.~\ref{fig:is} for the three filters. The red and grey colors
indicate the F and FO modes, respectively, and the two different
shading indicate models with masses 1.0-1.2 and 1.4 M$_\odot$. These regions become narrower and steeper moving towards
longer wavelengths (from F390W to F814W). Our observations are shown
for comparison and clearly define a very narrow region in period at
fixed luminosity, whereas the models span a larger period range. As
mentioned above, this is a fictitious consequence of using a linear
non adiabatic approach.
 To better constrain the models, the red boundary of
the IS is empirically fixed at {\it de-reddened} colors (F555W$-$F814W)$\sim$ 0.4
mag (see the black lines shown in Fig. \ref{fig:is}) which corresponds to the location in the CMD of the reddest SXPs \emph{observed} in our sample as well as in that of NGC5024 \citet{arellano11} with similar metallicity.

 To construct the pulsation
equations we therefore used only the models bluer than this limit and,
 owing to the limited range in colour, we also neglected the
 temperature dependence. Within these
assumptions, for each selected filter and pulsation mode, the grid of
available models allowed us to derive pulsation equations of the form
$\log$ M/M$_\odot = \alpha + \beta\times$MAG$+\gamma \log$ P$+\delta
\log$Z/Z$_\odot$, with the values of the coefficients listed in
Table~\ref{tab3}.

\begin{figure} 
\includegraphics[width=8.5cm]{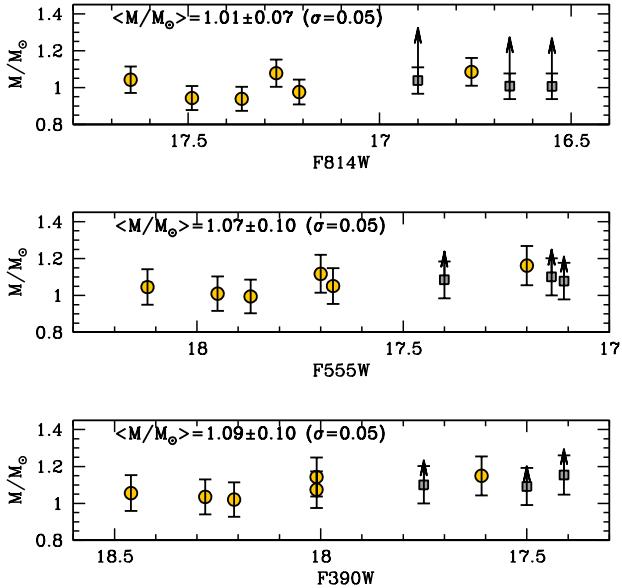} 
\caption{Pulsation masses as a function of magnitude for the nine SPX
  BSSs, estimated from the pulsation equations listed in Table
  \ref{tab3} (see Section \ref{sec:mass} for details). The mean values
  of the mass obtained in the three pass bands are labelled in each
  panel together with their error and their standard deviation. Symbols are the same
  used in Fig.~\ref{fig:pl}. Arrows point at the mass values derived when
F PL relations are used for stars with uncertain mode classification.}
\label{fig:masse}
\end{figure}

While the visual inspection of Fig.~\ref{fig:is} already suggests that
the observed variable BSSs have masses consistent with 1.0-1.2
M$_\odot$, the derived relations have been used to estimate the mean mass
and the dispersion for each star using the observational data in the
three pass bands. The results are listed in
Table~\ref{tab1} and summarized in Fig.~\ref{fig:masse},
where masses are plotted as a function of magnitudes in the three pass
bands and errorbars are estimated from the photometric uncertainty on
the mean magnitudes and the dispersion $\sigma$ of the
adopted relations (last column of the Table \ref{tab1}).  
In each panel we also give the mean mass value with its error
  computed on the whole sample, asssuming SXP2, SXP5 and SXP9 as FO pulsators. The standard deviation is also indicated to highlight the very good
stability of the data around the mean.

 Given the large uncertainty in the mode classification, we
have estimated the masses for SXP2, SXP5 and SXP9 also using the F
mode classification and the results are listed in Table~\ref{tab1}.
Moreover, the arrows in Fig.~\ref{fig:masse} start at the mass values
obtained assuming SXP2, SXP5 and SXP9 as FO pulsators and end at those
computed adopting the F mode classification. When we assume all
the stars as F pulsators, the mean mass values listed in each panel
of Fig.~\ref{fig:masse} slightly increase together with their standard
deviation, i.e. $<$M/M$_{\odot}$(F390W)$>$= 1.13$\pm$0.10 ($\sigma$=0.10), $<$M/M$_{\odot}$(F555W)$>$=
1.12$\pm$0.10($\sigma$=0.10) and $<$M/M$_{\odot}$(F814W)$>$=1.10$\pm$0.08($\sigma$=0.15).

\section{Discussion and Conclusions}
\label{discuss}

\begin{figure} 
\includegraphics[width=8.5cm]{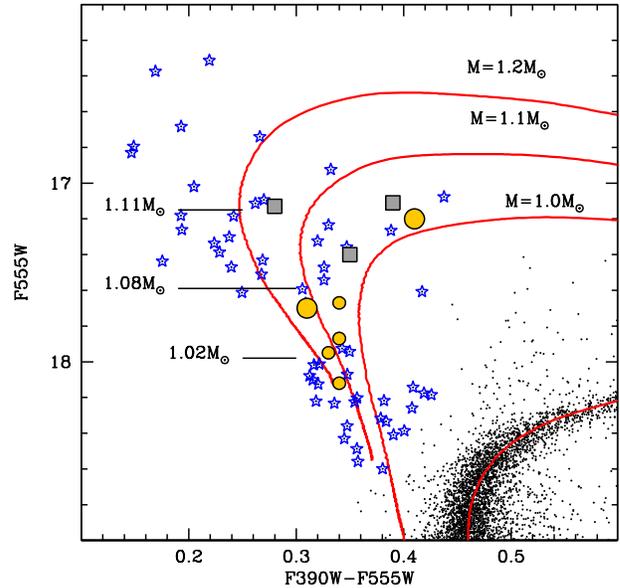} 
\caption{(F555W, F390W$-$F555W) CMD zoomed in the BSS region with
  highlighted the SXP BSSs (symbols as in Fig.~\ref{fig:pl}). The size
  of the symbols reflects the value of the estimated pulsation mass.
  The horizontal black segments indicate the mean masses (see labels)
  obtained by grouping the nine SXPs in three bins of increasing
  magnitude, and they are positioned in correspondence of the mean
  magnitude of each bin. Evolutionary tracks for single stars with
  masses equal to 1.0, 1.1 and 1.2 M$_\odot$
  \citep[from][]{pietrinferni04} are also shown for comparison.
 \label{fig:cmdmass}}
\end{figure}

We have used linear non adiabatic pulsation models \citep{santolamazza01} to estimate the mass
of the detected SXP BSSs as a function of their observed mean magnitude,
period and metallicity. As expected, we found that all the investigated SXP BSSs have
(pulsation) masses larger than the stellar mass at the MS-TO 
($\sim$0.75 M$_\odot$), as determined from $\alpha-$enhanced evolutionary
tracks at the cluster metallicity, computed from the BASTI database
\citep{pietrinferni04}. Within the errors, the obtained results
are compatible with an average mass $\langle$M/M$_{\odot}$$\rangle$ $\sim 1.06 \pm 0.09$ ($\sigma
=0.04$). However, Fig. \ref{fig:masse} shows a mild trend between mass and luminosity, especially if considering the two pass bands (F555W and
F390W) with the highest photometric accuracy (F814W, instead, is
significantly more affected by saturation problems). This is best
appreciable in Fig.~ \ref{fig:cmdmass}, where we have grouped the nine
SXPs in three magnitude bins and computed their average pulsation mass
(see the horizontal black segments and the corresponding labels in the
figure).This trend is even stronger when we classify the whole
  sample of SXPs as formed by F pulsators. In fact, the mean mass in
  each magnitude bin would be 1.01, 1.14 and 1.21 M$_{\odot}$ going from faint to
  brighter magnitudes.

In the same Fig.~\ref{fig:cmdmass}, the
position in the CMD of our SXP BSSs is compared to that of
$\alpha-$enhanced evolutionary tracks \citep{pietrinferni04} computed
for 1.0, 1.1 and 1.2 M$_\odot$ stars at the cluster metallicity and
assuming the same reddening and a distance modulus quoted above. 
These tracks bracket the location of our SXPs, thus suggesting a very
narrow range for the evolutionary mass, with a mean value $\langle$M$_{evo}\rangle$=1.1$\pm$0.1M${_\odot}$ which is in excellent agreement with the pulsation estimate.
A similar good agreement has been discussed by \citet{gilliland98}
using the pulsation properties of three mixed-mode SXPs in 47Tuc. These authors compared their observations in the
Petersen diagram with the same pulsation linear models used
here \citep{santolamazza01}. However, they also found a significant
discrepancy for one of the four analysed mixed-mode
SXPs that turns out to have a pulsation mass 20\% smaller than the
evolutionary one. 

In the last years, a variety of approaches have been used to estimate
the mass of BSS stars in globular and open
clusters. The first direct measurement of one of the most luminous
BSS in the core of 47 Tuc has been presented by \citet{shara97} through
a spectroscopic analysis of FOS@HST data. The derived mass
(1.7M$_{\odot}$) is twice the cluster's MS-TO mass and agrees well with
the estimate obtained using theoretical stellar evolutionary 
tracks for single star. 
In a subsequent study, using low- and intermediate-resolution
spectroscopy in a number of GCs (NGC6752, NGC5272 and NGC6397),
\citet{demarco05} found that the mean BSS masses 
for individual clusters (1.27, 1.05,
0.99, and 0.99M$_{\odot}$, respectively) were 
significantly lower than the values expected by evolutionary theory. However, this discrepancy can be accouted for by the large errors in their
mass evaluations. Only recently, very accurate mass estimates have been possible through spectroscopic and
photometric analysis of eclipsing binaries in three GCs, i.e. 47 Tucanae \citep{thompson10}, $\omega$ Centauri, NGC6752
\citep{kaluzny07a,kaluzny07b,kaluzny09} supporting the agreement between
these masses and those predicted from single star evolutionary
tracks. The only exception is BSS V209 in $\omega$ Centauri that shows a
larger mass ($\sim$0.95M$_{\odot}$) than MS-TO stars ($\sim$0.75M$_{\odot}$), but smaller
than what expected from the comparison between evolutionary tracks and its bright and
blue location in the CMD \citep[$\sim$1.2-1.3M$_{\odot}$, see][for more details]{kaluzny07a,ferraro06a}.
Finally, more recent observations devoted to the study of a
double-lined binary BSS in the not so young ($\sim$ 7 Gyr old) open
cluster NGC188 \citep{geller12} indicate that single star evolutionary models overestimate (by 15-30\%) the
dynamical mass.

While the reasons for some of the observed discrepancies need to be
clarified, the results here discussed indicate that single
star evolutionary tracks can be safely used to estimated the BSS mass
and suggest values below $\sim 1.2$ M$_\odot$ for most of the
population, in agreement with what argued in a number of previous
studies \citep[e.g.][]{ferraro06b,lanzoni07}. We also note
that, for BSSs close to the zero age MS, masses are more precisely
constrained by pulsation models, since stellar evolutionary tracks
become degenerate in that region of the CMD.

We conclude the paper by critically discussing the limitations of the
adopted approach. The models of \citet{santolamazza01} are suitable
for radial pulsators. Indeed, the relatively large amplitude of the
variations observed (A$_{\rm F555W} \gsim 0.05$ mag) and the period
ratio measured in double mode pulsators support the hypothesis that
most SXPs are radially pulsating stars \citep[as discussed
  in][]{stellingwerf79}. Moreover, despite a linear and non adiabatic
approach to the stellar pulsation, the models have been shown to well
reproduce the SXP periods and the blue boundary of the IS
\citep{stellingwerf79,gilliland98,santolamazza01}. In order to have a
full description of the radial stellar pulsation, non linear equations
have to be solved, including a treatment for the coupling between
pulsation and convection. Once available, these models will allow us
to address crucial open questions like the dependency of SXP pulsation
on masses, luminosities and chemical content, providing new tools to be used
to better constrain the pulsation masses of SXPs. We will have at
disposal more precise Period--Luminosity, Period--Luminosity--Color
relations and, more interestingly, we will be able to accurately predict the
amplitudes and the morphology of the SXP light curves. 
We have started the
computation of non linear models for SXP stars extending our
theoretical scenario to low masses and for different metallicities. As
described in \citet{bono02}, this requires very long computation time
and results will be presented in forthcoming papers, where we plan to apply
this new theoretical approach to all BSSs for which reliable mean
magnitudes and periods can be measured.

\section*{acknowledgments}
This research is part of the project Cosmic-Lab
(http://www.cosmic-lab.eu) funded by the European Research Council
under contract ERC-2010-AdG-267675.

\normalsize
\newpage


\begin{thebibliography}{68}
\expandafter\ifx\csname natexlab\endcsname\relax\def\natexlab#1{#1}\fi

\bibitem[{{Arellano Ferro} {et~al.}(2011){Arellano Ferro}, {Figuera Jaimes},
  {Giridhar}, {Bramich}, {Hern{\'a}ndez Santisteban}, \&
  {Kuppuswamy}}]{arellano11}
{Arellano Ferro}, A., {Figuera Jaimes}, R., {Giridhar}, S., {Bramich}, D.~M.,
  {Hern{\'a}ndez Santisteban}, J.~V., \& {Kuppuswamy}, K. 2011, \mnras, 416,
  2265

\bibitem[{{Bailyn}(1995)}]{bailyn95}
{Bailyn}, C.~D. 1995, \araa, 33, 133

\bibitem[{{Bellazzini} {et~al.}(1995){Bellazzini}, {Pasquali}, {Federici},
  {Ferraro}, \& {Pecci}}]{bellazzini95}
{Bellazzini}, M., {Pasquali}, A., {Federici}, L., {Ferraro}, F.~R., \& {Pecci},
  F.~F. 1995, \apj, 439, 687

\bibitem[{{Bono} {et~al.}(1997){Bono}, {Caputo}, {Cassisi}, {Castellani}, \&
  {Marconi}}]{bono97a}
{Bono}, G., {Caputo}, F., {Cassisi}, S., {Castellani}, V., \& {Marconi}, M.
  1997, \apj, 479, 279

\bibitem[{{Bono} {et~al.}(2002){Bono}, {Caputo}, {Marconi}, \&
  {Santolamazza}}]{bono02}
{Bono}, G., {Caputo}, F., {Marconi}, M., \& {Santolamazza}, P. 2002, in
  Astronomical Society of the Pacific Conference Series, Vol. 256,
  Observational Aspects of Pulsating B- and A Stars, ed. C.~{Sterken} \& D.~W.
  {Kurtz}, 249

\bibitem[{{Caputo} {et~al.}(2005){Caputo}, {Bono}, {Fiorentino}, {Marconi}, \&
  {Musella}}]{caputo05}
{Caputo}, F., {Bono}, G., {Fiorentino}, G., {Marconi}, M., \& {Musella}, I.
  2005, \apj, 629, 1021

\bibitem[{Cardelli {et~al.}(1989)Cardelli, Clayton, \& Mathis}]{cardelli89}
Cardelli, J., Clayton, G., \& Mathis, J. 1989, ApJ, 345, 245

\bibitem[{{Da Costa} {et~al.}(1986){Da Costa}, {Norris}, \&
  {Villumsen}}]{dacosta86}
{Da Costa}, G.~S., {Norris}, J., \& {Villumsen}, J.~V. 1986, \apj, 308, 743

\bibitem[{{Dalessandro} {et~al.}(2011){Dalessandro}, {Lanzoni}, {Beccari},
  {Sollima}, {Ferraro}, \& {Pasquato}}]{dalessandro11}
{Dalessandro}, E., {Lanzoni}, B., {Beccari}, G., {Sollima}, A., {Ferraro},
  F.~R., \& {Pasquato}, M. 2011, \apj, 743, 11

\bibitem[{{Dalessandro} {et~al.}(2008){Dalessandro}, {Lanzoni}, {Ferraro},
  {Vespe}, {Bellazzini}, \& {Rood}}]{dalessandro08}
{Dalessandro}, E., {Lanzoni}, B., {Ferraro}, F.~R., {Vespe}, F., {Bellazzini},
  M., \& {Rood}, R.~T. 2008, \apj, 681, 311

\bibitem[{{Dalessandro} {et~al.}(2013){Dalessandro}, {Ferraro}, {Massari},
  {Lanzoni}, {Miocchi}, {Beccari}, {Bellini}, {Sills}, {Sigurdsson},
  {Mucciarelli}, \& {Lovisi}}]{dalessandro13c}
{Dalessandro}, E., {et~al.} 2013, ArXiv e-prints

\bibitem[{{De Marco} {et~al.}(2005){De Marco}, {Shara}, {Zurek}, {Ouellette},
  {Lanz}, {Saffer}, \& {Sepinsky}}]{demarco05}
{De Marco}, O., {Shara}, M.~M., {Zurek}, D., {Ouellette}, J.~A., {Lanz}, T.,
  {Saffer}, R.~A., \& {Sepinsky}, J.~F. 2005, \apj, 632, 894

\bibitem[{{Dotter} {et~al.}(2010){Dotter}, {Sarajedini}, {Anderson},
  {Aparicio}, {Bedin}, {Chaboyer}, {Majewski}, {Mar{\'{\i}}n-Franch}, {Milone},
  {Paust}, {Piotto}, {Reid}, {Rosenberg}, \& {Siegel}}]{dotter10}
{Dotter}, A., {et~al.} 2010, \apj, 708, 698

\bibitem[{{Ferraro} {et~al.}(1991){Ferraro}, {Clementini}, {Fusi Pecci}, \&
  {Buonanno}}]{ferraro91}
{Ferraro}, F.~R., {Clementini}, G., {Fusi Pecci}, F., \& {Buonanno}, R. 1991,
  \mnras, 252, 357

\bibitem[{{Ferraro} {et~al.}(2001){Ferraro}, {D'Amico}, {Possenti}, {Mignani},
  \& {Paltrinieri}}]{ferraro01}
{Ferraro}, F.~R., {D'Amico}, N., {Possenti}, A., {Mignani}, R.~P., \&
  {Paltrinieri}, B. 2001, \apj, 561, 337

\bibitem[{{Ferraro} {et~al.}(1995){Ferraro}, {Fusi Pecci}, \&
  {Bellazzini}}]{ferraro95}
{Ferraro}, F.~R., {Fusi Pecci}, F., \& {Bellazzini}, M. 1995, \aap, 294, 80

\bibitem[{{Ferraro} {et~al.}(1992){Ferraro}, {Fusi Pecci}, \&
  {Buonanno}}]{ferraro92}
{Ferraro}, F.~R., {Fusi Pecci}, F., \& {Buonanno}, R. 1992, \mnras, 256, 376

\bibitem[{{Ferraro} {et~al.}(2006{\natexlab{a}}){Ferraro}, {Sollima}, {Rood},
  {Origlia}, {Pancino}, \& {Bellazzini}}]{ferraro06a}
{Ferraro}, F.~R., {Sollima}, A., {Rood}, R.~T., {Origlia}, L., {Pancino}, E.,
  \& {Bellazzini}, M. 2006{\natexlab{a}}, \apj, 638, 433

\bibitem[{{Ferraro} {et~al.}(2006{\natexlab{b}}){Ferraro}, {Sabbi}, {Gratton},
  {Piotto}, {Lanzoni}, {Carretta}, {Rood}, {Sills}, {Fusi Pecci}, {Moehler},
  {Beccari}, {Lucatello}, \& {Compagni}}]{ferraro06b}
{Ferraro}, F.~R., {et~al.} 2006{\natexlab{b}}, \apjl, 647, L53

\bibitem[{{Ferraro} {et~al.}(2009){Ferraro}, {Beccari}, {Dalessandro},
  {Lanzoni}, {Sills}, {Rood}, {Pecci}, {Karakas}, {Miocchi}, \&
  {Bovinelli}}]{ferraro09b}
---. 2009, \nat, 462, 1028

\bibitem[{{Ferraro} {et~al.}(2012){Ferraro}, {Lanzoni}, {Dalessandro},
  {Beccari}, {Pasquato}, {Miocchi}, {Rood}, {Sigurdsson}, {Sills}, {Vesperini},
  {Mapelli}, {Contreras}, {Sanna}, \& {Mucciarelli}}]{ferraro12}
---. 2012, \nat, 492, 393

\bibitem[{{Fiorentino} {et~al.}(2002){Fiorentino}, {Caputo}, {Marconi}, \&
  {Musella}}]{fiorentino02}
{Fiorentino}, G., {Caputo}, F., {Marconi}, M., \& {Musella}, I. 2002, \apj,
  576, 402

\bibitem[{{Fiorentino} {et~al.}(2006){Fiorentino}, {Limongi}, {Caputo}, \&
  {Marconi}}]{fiorentino06}
{Fiorentino}, G., {Limongi}, M., {Caputo}, F., \& {Marconi}, M. 2006, \aap,
  460, 155

\bibitem[{{Fiorentino} {et~al.}(2013){Fiorentino}, {Musella}, \&
  {Marconi}}]{fiorentino13a}
{Fiorentino}, G., {Musella}, I., \& {Marconi}, M. 2013, ArXiv e-prints

\bibitem[{{Fiorentino} {et~al.}(2010{\natexlab{a}}){Fiorentino}, {Contreras
  Ramos}, {Clementini}, {Marconi}, {Musella}, {Aloisi}, {Annibali}, {Saha},
  {Tosi}, \& {van der Marel}}]{fiorentino10a}
{Fiorentino}, G., {et~al.} 2010{\natexlab{a}}, \apj, 711, 808

\bibitem[{{Fiorentino} {et~al.}(2010{\natexlab{b}}){Fiorentino}, {Monachesi},
  {Trager}, {Lauer}, {Saha}, {Mighell}, {Freedman}, {Dressler}, {Grillmair}, \&
  {Tolstoy}}]{fiorentino10b}
---. 2010{\natexlab{b}}, \apj, 708, 817

\bibitem[{{Freire} {et~al.}(2008){Freire}, {Wolszczan}, {van den Berg}, \&
  {Hessels}}]{freire08}
{Freire}, P.~C.~C., {Wolszczan}, A., {van den Berg}, M., \& {Hessels}, J.~W.~T.
  2008, \apj, 679, 1433

\bibitem[{{Geller} \& {Mathieu}(2012)}]{geller12}
{Geller}, A.~M., \& {Mathieu}, R.~D. 2012, \aj, 144, 54

\bibitem[{{Gilliland} {et~al.}(1998){Gilliland}, {Bono}, {Edmonds}, {Caputo},
  {Cassisi}, {Petro}, {Saha}, \& {Shara}}]{gilliland98}
{Gilliland}, R.~L., {Bono}, G., {Edmonds}, P.~D., {Caputo}, F., {Cassisi}, S.,
  {Petro}, L.~D., {Saha}, A., \& {Shara}, M.~M. 1998, \apj, 507, 818

\bibitem[{{Harris}(1996)}]{harris96}
{Harris}, W.~E. 1996, \aj, 112, 1487

\bibitem[{{Hills} \& {Day}(1976)}]{hills76}
{Hills}, J.~G., \& {Day}, C.~A. 1976, \aplett, 17, 87

\bibitem[{{Jeon} {et~al.}(2003){Jeon}, {Lee}, {Kim}, \& {Lee}}]{jeon03}
{Jeon}, Y.-B., {Lee}, M.~G., {Kim}, S.-L., \& {Lee}, H. 2003, \aj, 125, 3165

\bibitem[{{Jeon} {et~al.}(2004){Jeon}, {Lee}, {Kim}, \& {Lee}}]{jeon04}
---. 2004, \aj, 128, 287

\bibitem[{{Jorgensen} \& {Hansen}(1984)}]{jorgensen84}
{Jorgensen}, H.~E., \& {Hansen}, L. 1984, \aap, 133, 165

\bibitem[{{Kaluzny} {et~al.}(2009){Kaluzny}, {Rozyczka}, {Thompson}, \&
  {Zloczewski}}]{kaluzny09}
{Kaluzny}, J., {Rozyczka}, M., {Thompson}, I.~B., \& {Zloczewski}, K. 2009,
  AcA, 59, 371

\bibitem[{{Kaluzny} {et~al.}(2007{\natexlab{a}}){Kaluzny}, {Rucinski},
  {Thompson}, {Pych}, \& {Krzeminski}}]{kaluzny07a}
{Kaluzny}, J., {Rucinski}, S.~M., {Thompson}, I.~B., {Pych}, W., \&
  {Krzeminski}, W. 2007{\natexlab{a}}, \aj, 133, 2457

\bibitem[{{Kaluzny} {et~al.}(2007{\natexlab{b}}){Kaluzny}, {Thompson},
  {Rucinski}, {Pych}, {Stachowski}, {Krzeminski}, \& {Burley}}]{kaluzny07b}
{Kaluzny}, J., {Thompson}, I.~B., {Rucinski}, S.~M., {Pych}, W., {Stachowski},
  G., {Krzeminski}, W., \& {Burley}, G.~S. 2007{\natexlab{b}}, \aj, 134, 541

\bibitem[{{Knigge} {et~al.}(2009){Knigge}, {Leigh}, \& {Sills}}]{knigge09}
{Knigge}, C., {Leigh}, N., \& {Sills}, A. 2009, \nat, 457, 288

\bibitem[{{Lanzoni} {et~al.}(2007){Lanzoni}, {Sanna}, {Ferraro}, {Valenti},
  {Beccari}, {Schiavon}, {Rood}, {Mapelli}, \& {Sigurdsson}}]{lanzoni07}
{Lanzoni}, B., {et~al.} 2007, \apj, 663, 1040

\bibitem[{{Lee} \& {Carney}(2002)}]{lee02}
{Lee}, J.-W., \& {Carney}, B.~W. 2002, \aj, 124, 1511

\bibitem[{{Leigh} {et~al.}(2011){Leigh}, {Sills}, \& {Knigge}}]{leigh11}
{Leigh}, N., {Sills}, A., \& {Knigge}, C. 2011, \mnras, 415, 3771

\bibitem[{{Leonard}(1989)}]{leonard89}
{Leonard}, P.~J.~T. 1989, \aj, 98, 217

\bibitem[{{Mapelli} {et~al.}(2007){Mapelli}, {Ripamonti}, {Tolstoy},
  {Sigurdsson}, {Irwin}, \& {Battaglia}}]{mapelli07}
{Mapelli}, M., {Ripamonti}, E., {Tolstoy}, E., {Sigurdsson}, S., {Irwin},
  M.~J., \& {Battaglia}, G. 2007, \mnras, 380, 1127

\bibitem[{{Marconi} {et~al.}(2004){Marconi}, {Fiorentino}, \&
  {Caputo}}]{marconi04}
{Marconi}, M., {Fiorentino}, G., \& {Caputo}, F. 2004, \aap, 417, 1101

\bibitem[{{Marconi} {et~al.}(2005){Marconi}, {Musella}, \&
  {Fiorentino}}]{marconi05}
{Marconi}, M., {Musella}, I., \& {Fiorentino}, G. 2005, \apj, 632, 590

\bibitem[{{Mateo} {et~al.}(1995){Mateo}, {Fischer}, \& {Krzeminski}}]{mateo95}
{Mateo}, M., {Fischer}, P., \& {Krzeminski}, W. 1995, \aj, 110, 2166

\bibitem[{{Mathieu} \& {Geller}(2009)}]{mathieu09}
{Mathieu}, R.~D., \& {Geller}, A.~M. 2009, \nat, 462, 1032

\bibitem[{{McCrea}(1964)}]{mccrea64}
{McCrea}, W.~H. 1964, \mnras, 128, 147

\bibitem[{{McNamara}(2011)}]{mcnamara11}
{McNamara}, D.~H. 2011, \aj, 142, 110

\bibitem[{{Monelli} {et~al.}(2012){Monelli}, {Cassisi}, {Mapelli}, {Bernard},
  {Aparicio}, {Skillman}, {Stetson}, {Gallart}, {Hidalgo}, {Mayer}, \&
  {Tolstoy}}]{monelli12a}
{Monelli}, M., {et~al.} 2012, \apj, 744, 157

\bibitem[{{Nemec} {et~al.}(1995){Nemec}, {Mateo}, {Burke}, \&
  {Olszewski}}]{nemec95}
{Nemec}, J.~M., {Mateo}, M., {Burke}, M., \& {Olszewski}, E.~W. 1995, \aj, 110,
  1186

\bibitem[{{Paresce} {et~al.}(1992){Paresce}, {de Marchi}, \&
  {Ferraro}}]{paresce92}
{Paresce}, F., {de Marchi}, G., \& {Ferraro}, F.~R. 1992, \nat, 360, 46

\bibitem[{{Petersen}(1978)}]{petersen78}
{Petersen}, J.~O. 1978, \aap, 62, 205

\bibitem[{{Pietrinferni} {et~al.}(2004){Pietrinferni}, {Cassisi}, {Salaris}, \&
  {Castelli}}]{pietrinferni04}
{Pietrinferni}, A., {Cassisi}, S., {Salaris}, M., \& {Castelli}, F. 2004, \apj,
  612, 168

\bibitem[{{Piotto} {et~al.}(2004){Piotto}, {De Angeli}, {King}, {Djorgovski},
  {Bono}, {Cassisi}, {Meylan}, {Recio-Blanco}, {Rich}, \& {Davies}}]{piotto04}
{Piotto}, G., {et~al.} 2004, \apjl, 604, L109

\bibitem[{{Pooley} \& {Hut}(2006)}]{pooley06}
{Pooley}, D., \& {Hut}, P. 2006, \apjl, 646, L143

\bibitem[{{Pych} {et~al.}(2001){Pych}, {Kaluzny}, {Krzeminski},
  {Schwarzenberg-Czerny}, \& {Thompson}}]{pych01}
{Pych}, W., {Kaluzny}, J., {Krzeminski}, W., {Schwarzenberg-Czerny}, A., \&
  {Thompson}, I.~B. 2001, \aap, 367, 148

\bibitem[{{Ransom} {et~al.}(2005){Ransom}, {Hessels}, {Stairs}, {Freire},
  {Camilo}, {Kaspi}, \& {Kaplan}}]{ransom05}
{Ransom}, S.~M., {Hessels}, J.~W.~T., {Stairs}, I.~H., {Freire}, P.~C.~C.,
  {Camilo}, F., {Kaspi}, V.~M., \& {Kaplan}, D.~L. 2005, Science, 307, 892

\bibitem[{{Sandage}(1953)}]{sandage53}
{Sandage}, A.~R. 1953, \aj, 58, 61

\bibitem[{{Santolamazza} {et~al.}(2001){Santolamazza}, {Marconi}, {Bono},
  {Caputo}, {Cassisi}, \& {Gilliland}}]{santolamazza01}
{Santolamazza}, P., {Marconi}, M., {Bono}, G., {Caputo}, F., {Cassisi}, S., \&
  {Gilliland}, R.~L. 2001, \apj, 554, 1124

\bibitem[{{Shara} {et~al.}(1997){Shara}, {Saffer}, \& {Livio}}]{shara97}
{Shara}, M.~M., {Saffer}, R.~A., \& {Livio}, M. 1997, \apjl, 489, L59

\bibitem[{{Stellingwerf}(1978)}]{stellingwerf78}
{Stellingwerf}, R.~F. 1978, \apj, 224, 953

\bibitem[{{Stellingwerf}(1979)}]{stellingwerf79}
---. 1979, \apj, 227, 935

\bibitem[{{Stetson}(1987)}]{stetson87}
{Stetson}, P.~B. 1987, \pasp, 99, 191

\bibitem[{{Stetson}(1994)}]{stetson94}
---. 1994, \pasp, 106, 250

\bibitem[{{Thompson} {et~al.}(2010){Thompson}, {Kaluzny}, {Rucinski},
  {Krzeminski}, {Pych}, {Dotter}, \& {Burley}}]{thompson10}
{Thompson}, I.~B., {Kaluzny}, J., {Rucinski}, S.~M., {Krzeminski}, W., {Pych},
  W., {Dotter}, A., \& {Burley}, G.~S. 2010, \aj, 139, 329

\bibitem[{{Vivas} \& {Mateo}(2013)}]{vivas13}
{Vivas}, A.~K., \& {Mateo}, M. 2013, ArXiv e-prints

\bibitem[{{Zinn} \& {Searle}(1976)}]{zinn76}
{Zinn}, R., \& {Searle}, L. 1976, \apj, 209, 734

\end{thebibliography}

\end{document}